\documentclass[%
 aps,%
 amsmath,amssymb,
 reprint,%
]{revtex4-1}

\usepackage{graphicx}
\usepackage{dcolumn}
\usepackage{bm}
\usepackage[inline]{enumitem}
\usepackage{natbib}

\begin{document}

\title[Modeling our survival in a zombie apocalypse]{Modeling our survival in a zombie apocalypse}

\author{Jo\~ ao Paulo A. de Mendon\c ca}
 \email{jpalastus@gmail.com}
 \affiliation{Dep. de F\' isica, Universidade Federal de Juiz de Fora, Brazil}
 \author{Lohan R. N. Ferreira}
  \affiliation{Dep. de Ci\^ encia da Computa\c c\~ ao, Universidade Federal de Juiz de Fora, Brazil}
\author{Leonardo M. V. Teixeira}
 \affiliation{Dep. de F\' isica, Universidade Federal de Juiz de Fora, Brazil}
\author{Fernando Sato}
 \affiliation{Dep. de F\' isica, Universidade Federal de Juiz de Fora, Brazil}

\date{\today}

\begin{abstract}
In this work we applied several concepts on the modeling of complex systems in an attempt to save mankind in the case of a zombie apocalypse. We developed a dynamical system in order to model a zombie outbreak in which we, regular humans, are aided by military personnel in the fight against the zombies. Our analysis has shown that the initial amount of military personnel play a key rule on our survival, even when the zombies are extremely aggressive and in large advantage.This document is a pre-print version of the paper \textit{de Mendonça, J.P.A., Teixeira, L.M.V., Sato, F. et al. Math Intelligencer (2019). https://doi.org/10.1007/s00283-019-09893-9}.
\end{abstract}

\pacs{
89.65.Ef;%Complex Systems
02.60.Cb;%Numerical simulation; solution of equations
05.70.Fh;
%Phase transitions: general studies (see also 05.30.Rt Quantum phase transitions in quantum statistical mechanics; 64.70.Tg Quantum phase transitions in specific phase transitions; 73.43.Nq Quantum phase transitions in quantum Hall effects; for superconductivity phase diagrams, see 74.25.Dw; for magnetic phase boundaries, see 75.30.Kz; for ferroelectric phase transitions, see 77.80.B-)
89.60.Gg
%Impact of natural and man-made disasters (for volcanic eruption effects on the atmosphere, see 92.60.Zc; for landslides, see 92.40.Ha; for floods, see 92.40.qp in Geophysics Appendix; for avalanches, see 92.40.vw in Geophysics Appendix; for global warming, see 92.70.Mn, for sea level change, see 92.70.Jw; for cryospheric change, see 92.70.Ha)
}

\keywords{Dynamical Systems; Numerical Modeling; Zombies; Military}
                              
\maketitle

\section{Introduction}

Here's the scenario: In a big city, on a very busy week day, a person who is supposedly sick turns into the first zombie specimen. From this point, the disease quickly spreads and soon the city becomes an apocalyptic scene, and we have no choice but to fight for our survival.

It may be too fantastic to seem real, but we've all seen it happen somewhere before, be it in a movie, on a computer game or our favorite science-fiction show. On those universes we, humans, are usually at a great disadvantage compared to the zombies, who usually are in greater numbers and multiply at the cost of our deaths.

One can find good works on literature applying mathematics and physics to many exotic systems, like soccer games\cite{davids2005} or in establishing romantic relationships\cite{fry2015}. We can also find a couple of other works on the modeling of zombies\cite{munz2009,hebert2014,badham2014}, but they usually rely in the confront between standard humans against the zombies, lacking on the use of our technological prowess and political advantages of living in a society to survive.

One could argue that zombies are fictional beings, and as such there is no point in modeling an outbreak, but even if a zombie apocalypse turns out to be something that happens only on the screens, learning how to mathematically model this kind of scenario can be very useful, as we can observe relations like this one in many other fields of research, such as economy \cite{edwards2017prey,mendoza2014austerity}, biology\cite{hughes2011behavioral,pantastico1992population} and social behavior\cite{jin2013epidemiological,abrams2011dynamics}. In all of those cases, the physics and mathematics behind the processes is the same, meaning that those systems can be modeled using the same set of equations. This means that any advance in one of those fields can help in the development of the others, by exploring their regent laws, knowing that they are related.

In this work, we build a physical model to describe the scenario of a zombie outbreak. This model diverges from the ones previously mentioned by introducing an extra population, comprised of military personnel. By numerical calculations of the time evolution of the populations, we analyzed the main behavior of the proposed system of equations for the four sets of parameters proposed and what they reveal about the zombie apocalypse outcome. 

\section{The Model}

We seek to create a model in which the humans could survive. In a human versus zombie competition, without external intervention, humans are usually annihilated\cite{cartmel2014}. To change this outcome we include a new group of individuals that are able to fight the zombies using weapons, whom we shall call military personnel. We then define $H(t)$, $Z(t)$ and $M(t)$ as the human, zombie and military personnel populations $t$ hours after the start of the zombie outbreak. The inclusion of the function $M(t)$ is the new feature we introduced, which is inexistent in the other models shown in literature. 

We can expect to see the three groups competing for space in the city. To model this space competition between the groups, we assume the city to be composed of discretized cells, of which $S$ are available to the individuals(like cells on a discretization, or a big 3D game board). With this model, the average number of encounters per time unit ($N_{AB}$) of two randomly distributed populations, $A$ and $B$, will be given by:

\begin{equation}
\begin{split}
N_{AB} &= S \times P(\text{A} \land \text{B})\\
%&= S \times P(\text{A})\times P(\text{B})\\
%&=S\times \frac{A(t)}{S} \times \frac{B(t)}{S}\\
&=\frac{A(t) \times B(t)}{S}\\
\end{split}
\end{equation}

Where we denote $P(\text{A})$ as the probability of finding a member of group A on a given cell. Notice that $N_ {AB}$ is a function of time. Thus, we are interested in the values $N_{HZ} = H(t)Z(t)S^{-1}$, $N_{MZ} = M(t)Z(t)S^{-1}$ and $N_{MH} = M(t)H(t)S^{-1}$.

Now we need to establish what happens when individuals from two different populations meet. When a zombie encounters a human, we can imagine three possible scenarios: (i) The human kills the zombie, (ii) the zombie kills the human or (iii) the human is zombified (i.e., turns into a zombie). For simplicity, let's consider these three events to be mutually exclusive, and that the probability that any of the three occurs equals 1, that is, when a human-zombie encounter occurs, one of those things will certainly happen, and only one. In order to create a pattern and simplify our notation, we will call $\alpha _{HZ}$ the probability of event (i), $\beta _{HZ}$ the probability of event (ii) and $ \kappa _{HZ} $ the probability of event (iii) happening. Then the above relation can be written as:

\begin{equation}
\alpha _{HZ} + \beta _{HZ} + \kappa _{HZ} = 1 
\label{somaHZ}
\end{equation}

The case where a soldier(military personnel) encounters a zombie has a very similar logic, but the soldier must have greater ease in killing the zombies, since they are trained and armed. It can also be expected that the probability of a soldier turning into a zombie has to be lower than that of a human turning into a zombie, since they are trained to defend themselves and have greater physical stamina and constitution. We then define the new parameters $\alpha _{MZ}$, $\beta _{MZ}$ and $\kappa _{MZ}$ in the same sense as we did for the human-zombie case, but with different values in order to adjust for them being in the military. Again, we can write the relation:

\begin{equation}
\alpha _{MZ} + \beta _{MZ} + \kappa _{MZ} = 1 
\label{somaMZ}
\end{equation}

At last, in the case where a human encounters a military personnel, since they are on the same side, we'd expect no death outcomes, but instead it may happen that, in order to increase their overall survival chances, the military chooses to arm or train that human individual, turning him into a soldier. Therefore, we define the value $\kappa _{HM}$ as the probability that, in one of those encounters, a human becomes a military personnel.

With all the relevant variables defined, we are ready to model the equations that describe the time evolution of the system. In order to do the time evolution, we assume that we know $H(t)$, $Z(t)$ and $M(t)$ at a given time $t$. We choose to first process the encounters between humans and zombies, then calculate the resulting populations after these encounters, on average. If we name $t'$ the time right after all such encounters, then the final populations are given by the equations:

\begin{equation}
  \begin{cases}
    H(t') &= H(t) - N_{HZ}(\kappa _{HZ} + \beta _{HZ})\\
    Z(t') &= Z(t) + N_{HZ}(\kappa _{HZ} - \alpha _{HZ})\\
    M(t') &= M(t)\\
  \end{cases} 
\end{equation}

Note that, in the above equations, $N_{HZ}\kappa _{HZ}$ is the average number of humans turned into zombies, $N_{HZ}\alpha _{HZ}$ that of zombies killed and $N_{HZ}\beta _{HZ}$ that of humans killed. Next, we consider the encounters between the military and zombies. For these encounters, the relevant system of equations will be similar to the ones from the previous case:

\begin{equation}
  \begin{cases}
    H(t'') &= H(t') \\
    Z(t'') &= Z(t') + N_{MZ}(\kappa _{MZ} - \alpha _{MZ})\\
    M(t'') &= M(t') - N_{MZ}(\kappa _{MZ} + \beta _{MZ})\\
  \end{cases} 
\end{equation}

At last, for the encounters between a human and the military, we have the following equations:

\begin{equation}
  \begin{cases}
    H(t+1) &= H(t'') - N_{HM}\kappa _{HM}\\
    Z(t+1) &= Z(t'') \\
    M(t+1) &= M(t'') + N_{HM}\kappa _{HM}\\
  \end{cases} 
\end{equation}

Where $t+1$ denotes the moment after all encounters. The order of the encounters has been chosen to be this one simply in order to follow the logic presented in this paper. We see that the bigger the value of $S$ is compared to the initial populations, the lesser the choice we make for this order will matter. This process can then be repeated, generating a series of values for $H$, $M$ and $Z$, which evolve as $t$ increases. This is what defines $t$ as our temporal variable and this process as the time evolution of the populations.

With the steps that need to be taken made clear, it's time to write the algorithm to run the numerical calculations. Thinking about computation time, an easier implementation and future adaptations, we use an algorithm (available as Supplementary Material) developed in Java with BlueJ (a free Java Development Environment designed for beginners and more advanced users).

In order to calculate the temporal evolution of our dynamic variables, we defined the numerical values for the parameters as follows:
\begin{enumerate*}[label=(\roman*)]
\item The city has an initial population of 1000 humans, i.e., $H(t=0)=1000$. 
\item The system starts with only one zombie ($Z(t=0)=1$). The important  factor here is the initial ratio  between $Z$ and $H$, that is equal to $0.001$ in this case. Different values for this ratio can change the simulation's outcome.    
\item As the humans' success depends on the performance of the military, we will vary the value of $M(t=0)$ and try to find if there exists a minimum initial number of military personnel needed so that we can survive. 
\item The value of $S$ is set to 2000. Notice that $S$ determines the rhythm in which the system will evolve, so its value is arbitrary. Bigger values can be used in order to make the system evolve more slowly and reduce the effect of the order in which the encounters occur. However, large values of $S$ can result in very time-consuming calculations.
\item Our stop conditions are: The joint population of humans and military personnel becomes less than $1$ (we call this outcome a failure) or the number of zombies becomes less than $0.9$ (just to be sure. We call this outcome a success.).  
\end{enumerate*}

As for the parameters $\alpha$, $\beta$ and $\kappa$, we present four sets of values representing four different situations.

\subsection*{Parameters for a Strong Military}
 
This is our standard model. When humans encounter zombies, we set 60\% as the probability that the human dies, 10\% as the probability that the zombie dies and 30\% as the probability for zombification. As the military personnel are more trained than the regular humans, and we are assuming that they are strong enough to fend for themselves, we defined that in 10\% of the cases a soldier dies and in 10\% it is zombified. On the remaining 80\% of the encounters, the soldiers kill the zombies and take a step forwards in the direction of our salvation. As not every human has the aptitude to become a soldier, and such training takes time, we opted to set the chance that a human is trained to become a soldier to be only 1\%. A summary of these parameters can be seen in table \ref{tab:mf}.

\begin{table}[htbp]
\centering
\small
\begin{tabular}{|l|l|l|l|}
\hline
\textbf{} & {$\beta$} & {$\alpha$} & {$\kappa$} \\ \hline
\textbf{HZ} & 0.6 & 0.1 & 0.3 \\ \hline
\textbf{MZ} & 0.1 & 0.8 & 0.1 \\ \hline
\textbf{HM} & - & - & 0.01 \\ \hline
\end{tabular}
\caption{Parameters for the case of a strong military.}
\label{tab:mf}
\end{table}

\subsection*{Parameters for Violent Zombies}

These parameters represent the case where we have zombies with high levels of violence. Such zombies are more likely to kill their prey, so the numerical values of $\beta_{HZ}$ and $\beta_{MZ}$ were reevaluated to $0.8$ and $0.34$, respectively. Then, considering the relations given in equations \ref{somaHZ} and \ref{somaMZ}, we adapt the remaining parameters in order to keep the notion of a probability. The parameters we used for this model are summarized in table \ref{tab:sz}.

\begin{table}[htbp]
\centering
\small
\begin{tabular}{|l|l|l|l|}
\hline
\textbf{} & {$\beta$} & {$\alpha$} & {$\kappa$} \\ \hline
\textbf{HZ} & 0.8 & 0.05 & 0.15 \\ \hline
\textbf{MZ} & 0.34 & 0.59 & 0.07 \\ \hline
\textbf{HM} & - & - & 0.01 \\ \hline
\end{tabular}
\caption{Parameters for the case of violent zombies.}
\label{tab:sz}
\end{table}

\subsection*{Parameters for a Very Strong Military}

This is also a modification of our first model. We changed the values of $\beta _{MZ}$, $\alpha _{MZ}$ and $\kappa _{MZ}$ to $0.015$, $0.98$ and $0.005$. These values have been chosen in order to consider extremely well-trained elite soldiers, capable of killing zombies in 98\% of the encounters. As they are special military forces, we also reduce the probability that a human can be trained into one of them to 0.1\%. The final set of parameters is then shown in table \ref{tab:mmf}.

\begin{table}[htbp]
\centering
\small
\begin{tabular}{|l|l|l|l|}
\hline
\textbf{} & {$\beta$} & {$\alpha$} & {$\kappa$} \\ \hline
\textbf{HZ} & 0.6 & 0.1 & 0.3 \\ \hline
\textbf{MZ} & 0.015 & 0.98 & 0.005 \\ \hline
\textbf{HM} & - & - & 0.001 \\ \hline
\end{tabular}
\caption{Parameters for the case of a very strong military.}
\label{tab:mmf}
\end{table}

\subsection*{Parameters for Resistant Humans}

In this last set of parameters, we choose to describe a special condition in which humans are healthier and stronger, being more capable to fend for their lives and kill zombies. We set the values as $\beta _{HZ}=0.3$, $\alpha _{HZ}=0.3$ and $\kappa _{HZ}=0.4$. Notice that the added resistance makes those humans less likely to be eaten, but the chance of one turning into a zombie is a little greater(as body endurance is not enough to resist the pathogen). Also, we set $\kappa _{HM}=0.1$, because such regular humans can now be more easily trained  to become a soldier. All these values are shown in table \ref{tab:hr}.

\begin{table}[htbp]
\centering
\small
\begin{tabular}{|l|l|l|l|}
\hline
\textbf{} & {$\beta$} & {$\alpha$} & {$\kappa$} \\ \hline
\textbf{HZ} & 0.3 & 0.3 & 0.4 \\ \hline
\textbf{MZ} & 0.1 & 0.8 & 0.1 \\ \hline
\textbf{HM} & - & - & 0.1 \\ \hline
\end{tabular}
\caption{Parameters for the case of resistant humans.}
\label{tab:hr}
\end{table}

\section{Results and analysis}

Our system reveals itself to be very dependent on the parameters we choose and the initial values of the populations. We have done a series of calculations for each set of parameters, keeping the initial number of humans and zombies fixed and changing the initial number of military personnel ($M(t=0)$). Let us discuss the major results obtained.

\begin{figure}[htbp]
    \centering
    \includegraphics[height=5cm]{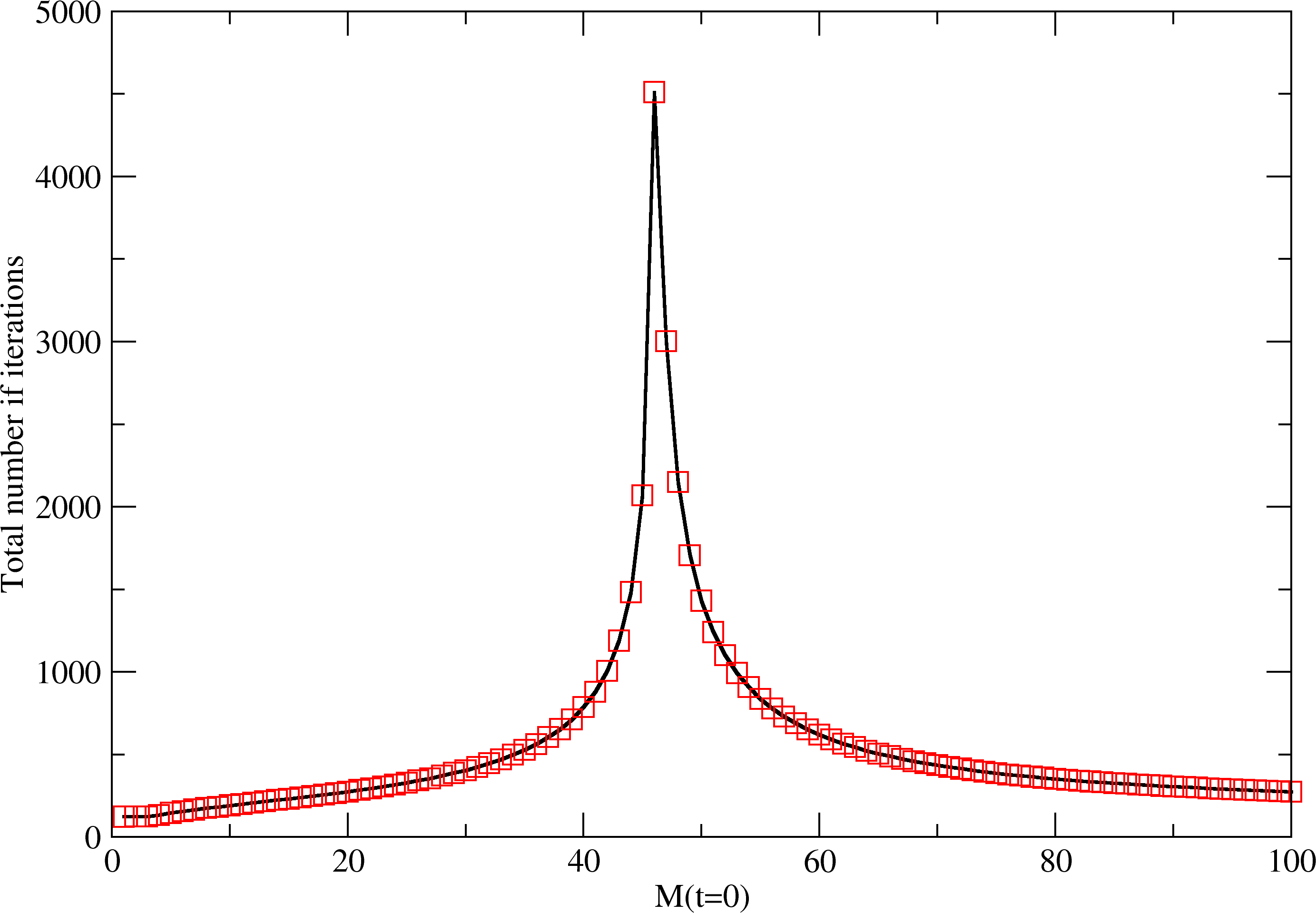}
    \caption{Plot showing the relation between the initial number of military personnel and the total number of iterations, i.e., the $t$ where the algorithm finds our stopping condition to be true. These results are for the parameters of the Strong Military case. The peak in $M(t=0)=46$ indicates a critical point, which characterizes the phase transition present in our model.}
    \label{fig:mf_criticalpoint}
\end{figure}

First, for some sets of parameters, our system shows a phase transition, which can be seen in figure \ref{fig:mf_criticalpoint}. The points to the left of the peak represent the dynamics where the zombies are the winners. In this case, the bigger the initial number of military personnel is, the longer is the time that the zombies spend fighting with them in order to end the battle. This is reflected on the inclination of the curve, which is always crescent before the critical point. Getting to the critical point, the time that the zombies spend killing the humans becomes so large that the military ends up killing a large amount of zombies, and changing the outcome of the war. The left side can be seen as a phase in which the zombies win, whereas the right side is a phase in which the humans win. In this sense, the highest point in this graph characterizes a phase transition.  

\begin{figure}[htbp]
    \centering
    \includegraphics[height=5cm]{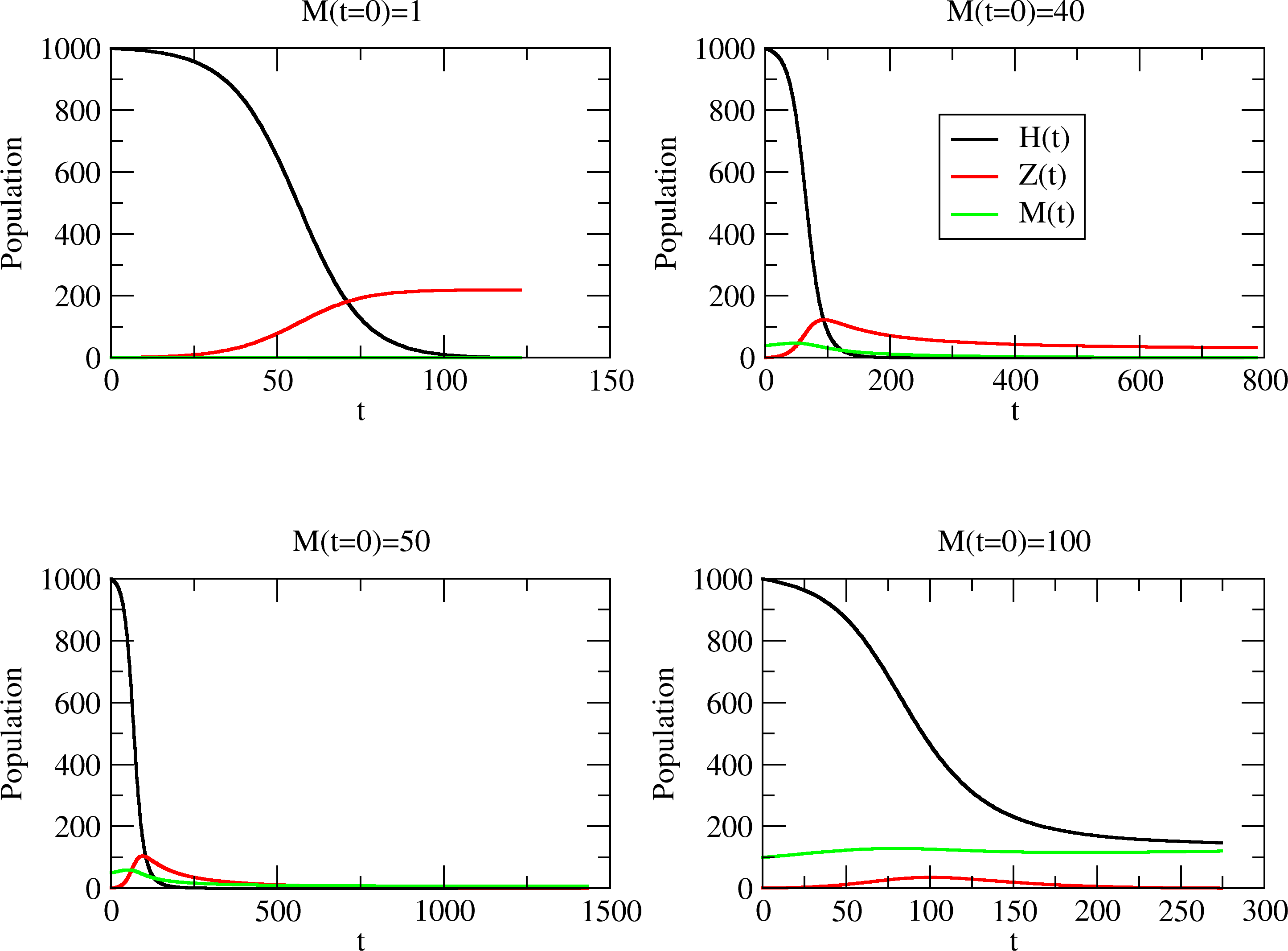}
    \caption{These plots, for the parameters of the Strong Military case, are examples of the system's evolution over time for different values of $M(t=0)$.}
    \label{fig:mf_evolution}
\end{figure}

In figure \ref{fig:mf_evolution} we show in more detail what happens in each phase. For $M(t=0)=0$, we have our worst situation. In just a few interactions the population of humans becomes smaller than $1$. For $M(t=0)=40$, the number of soldiers increases at the start of the simulation, even though this growth is not enough to stop the zombies. In this simulation, it is also possible to observe the population of zombies starting to decline after $t\approx 100$. The value $M(t=0)=46$ is in fact the last simulation in this series that represents an evolution in the zombie-victory phase. Nonetheless, the graph for $M(t=0)=50$ shows only a few human survivors. Better scenarios came with the increase on our initial military population. For $M(t=0)=100$, we end up with a survival of approximately 15\% of our initial population.    

\begin{figure}[htbp]
    \centering
    \includegraphics[height=5cm]{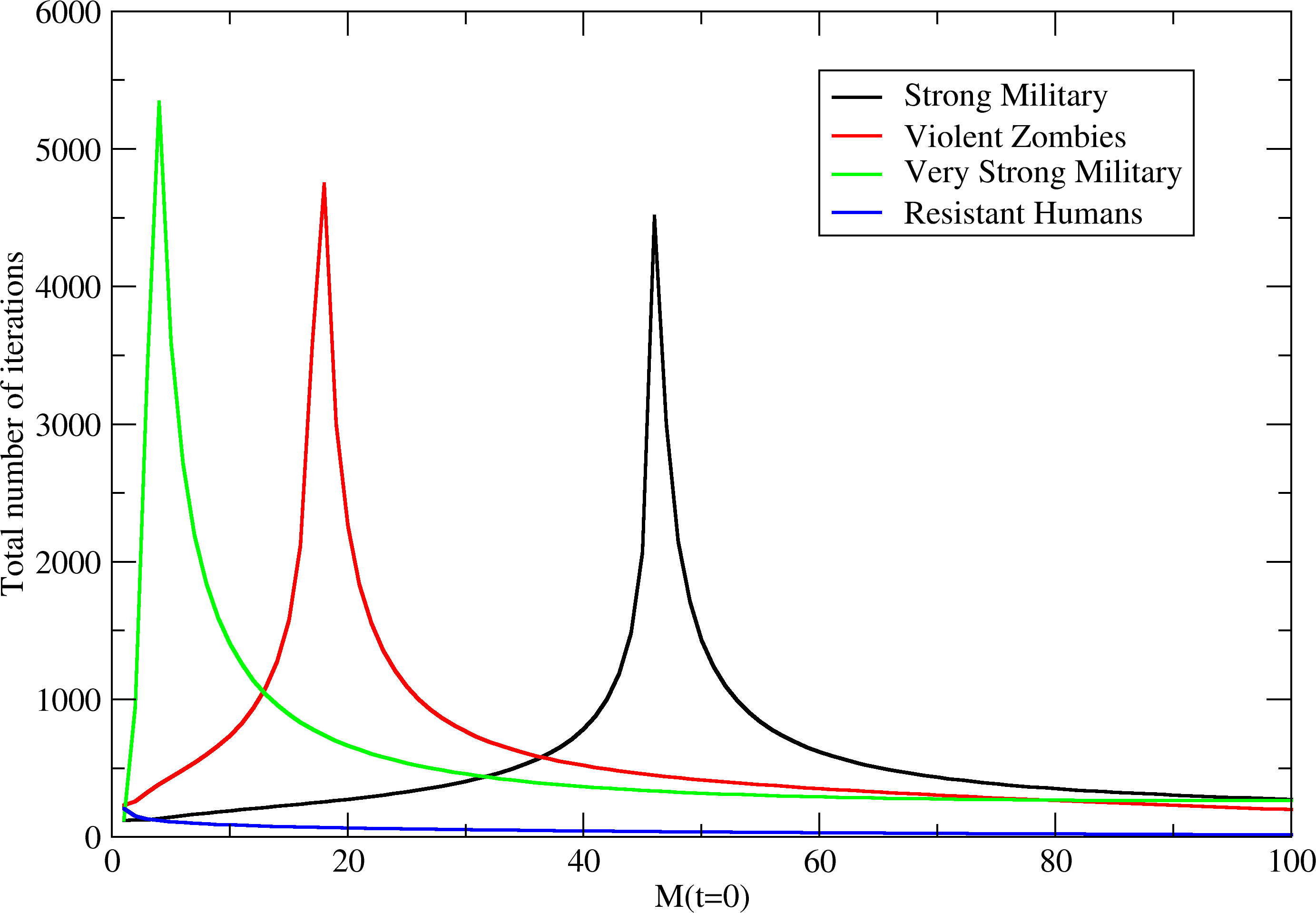}
    \caption{The different values chosen for our models change the value of $M(t=0)$ where the critical point occurs. Here, we show the correspondent of Figure \ref{fig:mf_criticalpoint} for all of our proposed sets of parameters.}
    \label{fig:criticalpoints}
\end{figure}

Figure \ref{fig:criticalpoints} shows how the plot of figure \ref{fig:mf_criticalpoint} changes for all of our parametrizations.   

First, we see that for the case of Violent Zombies we need a lesser amount of military personnel than in our standard model. This surprising result comes from the fact that these zombies are so efficient in killing humans that their population doesn't increase in time as strongly as in the other cases. The down side of this model is that the population of regular humans quickly disappears, and the only survivors are soldiers. 

The critical point also appears at smaller values of $M(t=0)$ for the model of Very Strong Military. An impressive result comes with the Resistant Humans model, in which we didn't have a critical point. With these parameters, all the simulations resulted in a large advantage of the humans over the zombies. The major reason why these results are generated is that the humans could rapidly be trained as military personnel and suppress the zombie invasion more efficiently. 

\section{Conclusions}

Our objective was to figure out what would happen if a zombie outbreak were to start out of the blue. For such, we created a theoretical model for this system and thereby figured out what happened in this simulated reality. This model, which in essence is a dynamic system\cite{devaney1993,temam2012infinite}, has parameters that are directly related to the probabilities of the outcomes of the encounters between survivors and zombies. As doing a thousand experiments by putting a human and a zombie inside of a box is out of question, we chose hypothetical values for those parameters by thinking of four possible scenarios: A standard model with Strong Military, a model with Violent Zombies, one with Very Strong Military and other with Resistant Humans.

Our results indicate that the military intervention is fundamental to the survival of mankind. Their presence introduces a critical point in our dynamics, which represents the separation between two phases: Survive or be eaten. Also, this phase transition is characterized by a peak in the graph shown in figure \ref{fig:mf_criticalpoint}, and it reflects the fact that having just the minimum number of military personnel necessary to win will make the fight last longer. 

From our standard model, with strong military, we were able to conclude that a number of 47 soldiers per 1000 capta is the minimum number necessary to human survival. Comparing this numbers with data from \textit{The Military Balance 2017}\cite{iiss2017military}, North Korea has enough military personnel in activity (47.4 per 1000 capta) to survive our modeled zombie outbreak. On the other hand, the calculations with the Strong Military parameters ended up revealing a terrible decrease in the civilian population, resulting in a survival of only 12.1\% of the initial civil population, with $M(t=0)=100$ (the best case studied for this model). 

However, we have seen that the model with resistant humans is incredibly promising, since all of the simulations for these parameters have shown a victory of mankind. As this case represents a reality that's very different from our standard model, we see that humanity can evolve until it becomes able to handle a zombie outbreak, with the individuals themselves being stronger and more capable to fight for their own survival. Just to compare, with $M(t=0)=100$, this model reveals that 21.0\% of the initial population of resistant humans survives the outbreak. So we conclude that, be it for North Korea or for other countries like the United States (4.2 active military per 1000 capta) or Brazil (1.6 active military per 1000 capta)\cite{iiss2017military}, the best strategy to save mankind, suggested by our model, is to invest in their own populace, making their standard humans more healthy and more capable to survive.

Now, if you're not a politician or have come to this paper just because of the zombies, here's some advice: Take some of the candy out of your diet, go practice some physical exercises and sports. Perhaps a martial art? According to our calculations, this is the best way to ensure that we are ready when the survival of our species depends on us.

\section*{Acknowledgements}

We would like to thank Prof. Dr. Rodrigo Luiz de Souza da Silva and Prof. Dr. Ilya Lvovich Shapiro, for all the direct and indirect support. We acknowledge the Brazilian Agencies CNPq, CAPES, FAPEMIG and FINEP for the financial support of this work.

\end{document}